\documentclass{ws-ijqi}

\usepackage{graphics}
\usepackage{amsmath}
\usepackage{amstext}
\usepackage{latexsym}
\usepackage{graphicx}
\usepackage{epstopdf}
\usepackage{color}
\usepackage{subfigure}
\usepackage{epsfig}

\begin{document}
\let\veps=\varepsilon
\let\vphi=\varphi
\let\de=\partial
\def\I {{\rm 1} \hspace{-1.1mm} {\rm I} \hspace{0.5mm}}
\def\Z {{\sf Z\hspace{-1ex}Z}}
\def\R {{ \sf I\hspace{-.2ex}{\sf R}}}
\def\C {{\sf I\hspace{-1.3ex}{\sf C}}}
\def\N {{\sf I\hspace{-.2ex}{\sf N}}}
\def\M {{\sf I\hspace{-.2ex}{\sf M}}}
\def\re{{ \rm I\hspace{-.3ex}{\rm R}}\mbox{e}}
\def\im{{\rm I} \hspace{-0.9mm} {\rm I} \hspace{0.1mm}\mbox{m}}
\newcommand\ket[1]{\left|#1\right\rangle}
\newcommand\bra[1]{\left\langle#1\right|}
\newcommand\braket[2]{\left.\left\langle#1\right|#2\right\rangle}

\newcommand\be{\begin{equation}}
\newcommand\ee{\end{equation}}

\markboth{F. Plastina and S. Maniscalco} {Non-Markovian dynamics
of system-reservoir entanglement}

\catchline{}{}{}{}{}

\title{Non-Markovian dynamics of system-reservoir entanglement}

\author{F. PLASTINA}

\address{Dip. Fisica, Universit\`a della Calabria,
87036 Arcavacata di Rende (CS) Italy \\ INFN - Gruppo collegato di
Cosenza \\ francesco.plastina@fis.unical.it}

\author{S. MANISCALCO}

\address{Turku Center for Quantum
Physics, Department of Physics and Astronomy, University of Turku,
FIN-20014 Turku, Finland \\ SUPA, EPS/Physics, Heriot-Watt
University, Edinburgh, EH144AS, UK \\ s.maniscalco@hw.ac.uk }

\maketitle

\begin{history}
\received{Day Month Year} \revised{Day Month Year}
\end{history}

\begin{abstract}
Using an exact approach, we study the dynamics of entanglement
between two qubits coupled to independent reservoirs and between
the two, initially disentangled, reservoirs. We also describe the
transfer of bipartite entanglement from the two-qubits to their
respective reservoirs focussing on the case of two atoms inside
two different leaky cavities with a specific attention to the role
of the detuning. We present a scheme to prepare the cavity fields
in a maximally entangled state, without direct interaction between
the cavities, by exploiting the initial qubits entanglement. We
discuss a deterministic protocol, working in the presence of
cavity losses, for the generation of a W-state of one qubit and
two cavity fields and we describe a probabilistic scheme to
entangle one of the atoms with the reservoir (cavity field) of the
other atom.
\end{abstract}

\section{Introduction}
The decoherence of entangled systems has been the subject of an
intense research activity in the last few years, triggered by the
theoretical prediction of the phenomenon of entanglement sudden
death (ESD) \cite{yu}, experimentally demonstrated with entangled
photon pairs \cite{alme} and atomic ensembles \cite{laurat}.
Subsequent theoretical paper have extended the original work in
various directions, by studying the dynamics of entangled systems
coupled to either the same \cite{noi} or different reservoirs,
including non-markovian effects \cite{bellomo} experimentally
tested in \cite{cialdi}; by examining the effect of finite
temperature reservoirs \cite{alq}; by including counter-rotating
terms \cite{jing} and by exploring the role of diversity between
the subsystems \cite{schan,ferdi}. Furthermore, it has been
understood that the sudden death phenomenon is, in fact, rather a
transfer, as the environmental degrees of freedom become entangled
while the two subsystems disentangle \cite{lopez} (entanglement
sudden birth, ESB).

The study of reservoirs entanglement in relationship to the
phenomenon of entanglement sudden death has been also performed
for two-level-systems interacting with two ideal cavities
\cite{yudue,chen}. Further investigations have been devoted to the
existence of entanglement invariant \cite{mansei}.

The first aim of this paper is to study the sudden birth
phenomenon in the case of non-markovian reservoirs, in order to
elucidate the role of entanglement memory on entanglement
exchange. In doing this, we will show that the two cases of {\it
i})a spectrally flat, markovian environment, and of {\it ii})a
monochromatic environment, first studied in Refs. \cite{lopez} and
\cite{yudue}, respectively, are just two manifestation of the same
phenomenon, as both of them can be described in a unique fashion
(see also Ref. \cite{xu}). Indeed, the Hamiltonian model that we
discuss continuously interpolates between these two extrema
depending on the spectral width of the environmental projected
density of states.

More explicitly, we present an exact approach to the dynamics of
two qubits interacting with independent structured reservoirs
having Lorentzian spectral distributions. This system describes,
e.g., two atoms inside two leaky cavities, whose quality factors
are related to the widths of the Lorentzians, and provides a quite
realistic description of the case of two ions trapped into two
cavities, as discussed in Ref. \cite{harkonen}. The solution of
this model reduces to the Markovian one in the bad cavity limit
and to the Jaynes-Cummings one in the ideal cavity limit.

The exact solution of the total closed system allows to describe
the zero-temperature reservoirs as effective two-state systems,
therefore the entanglement between the qubits, between each qubit
and its own reservoir and between the two reservoirs can be
characterized by using the concurrence \cite{wootte}. In this
framework we show how, for the general non-Markovian case, the
disappearance of the initial atomic entanglement in a finite time
(ESD) is accompanied by the sudden appearance of entanglement in
the reservoirs (ESB), as already demonstrated in the Markovian
case, and discuss with particular attention the role played by the
detuning of the atoms with respect to the cavity fields. Moreover
we prove that the qubit-reservoir tangle is an entanglement
invariant.

Finally, we describe a deterministic scheme to manipulate the
system in the presence of losses, in order to generate a W-state
of one atom and two cavity fields. Starting from this state, and
for certain values of the system parameters, the subsequent
measurement of the logical state of the qubit projects the cavity
fields in a maximally entangled state. Moreover we demonstrate how
one of the two atoms can be maximally entangled with the field
inside the other cavity. Similarly to the case of entanglement
swapping, we show that quantum systems that have never interacted
directly, namely one atom and the field inside a distant cavity,
can be entangled if some entanglement is present in the initial
two-atom state. As already mentioned, these procedures work
independently of the amount of non-markovianity of the
environment; that is, independently of the quality factors of the
two cavities.

The paper is structured as follows. In Sec. \ref{sec:2} the
analytic solution of the model is presented together with and its
Markovian and ideal cavity limits; Sec. \ref{sec:3} shows the main
features of the entanglement dynamics; in Sec. \ref{sec:4} the
scheme for entangled state generation is reported, while a summary
and some final remarks are given in Sec. \ref{sec:5}.

\section{Exact dynamics}\label{sec:2}

Consider a pair of two-level systems initially entangled to some
extent and interacting with two different environments; a
situation realized, e.g., for two initially excited (and
entangled) atoms spontaneously emitting into two different
cavities. Each of the two atom-cavity systems is described by
Hamiltonian of the kind \be H^{(i)}_{\alpha} = \sigma^+_{\alpha}
\sum_{k_{\alpha}} g_{k_{\alpha}} b_{k_{\alpha}} e^{i
\delta_{k_{\alpha}} t} + \mbox{h. c.} \ee This Hamiltonian is
written in the interaction picture with respect to the free atom
and field terms. Here $\alpha=1,2$ distinguishes between the two
atom-cavity system, $\sigma_{\alpha}^{(\pm)}$ are the rising and
lowering operator for the atom $\alpha$ spontaneously emitting a
photon into the $k_{\alpha}$ modes, for which $b_{k_{\alpha}}$ are
the annihilation operators. Finally the detuning is defined as the
difference between the atomic and the mode frequencies
$\delta_{k_{\alpha}} = \omega_{\alpha} - \omega_{k_{\alpha}}$.

Each of the two sets of cavity modes is specified by its spectral
density and time correlation function:
$$f_{\alpha}(\tau) = \sum_{k_{\alpha}} |g_{k_{\alpha}}|^2 e^{i \delta_{k_{\alpha}}
\tau} \equiv \int d\omega J_{\alpha}(\omega) e^{i
\delta_{k_{\alpha}} \tau} \, .$$ We will assume that each cavity
has a finite photon lifetime and describe this by assuming
Lorentzian broadened spectral densities $$J_{\alpha}(\omega) =
\frac{\Omega^2_{\alpha}}{ \pi} \, \frac{\lambda_{\alpha}}{(\omega
- \omega_{c_{\alpha}})^2 + \lambda_{\alpha}^2} \, , $$ where
$\omega_{c_{\alpha}}$ is the central frequency of cavity $\alpha$,
while $\lambda_{\alpha}^{-1}$ is the cavity life-time. For future
reference, we also define the detuning of the atomic transition
from the frequency of the main cavity mode: $\delta_{\alpha} =
\omega_{\alpha}- \omega_{c_{\alpha}}$. As one can evaluate using
the residue method, such density of states leads to a correlation
function of the form
$$f(\tau) = \int_{-\infty}^{+\infty} d \omega \, J e^{i
(\omega_{c_{\alpha}}-\omega) \tau} = \frac{\gamma_{\alpha}
\lambda_{\alpha}}{2} e^{- \lambda_{\alpha} |\tau|} \equiv
\Omega_{\alpha}^2 \, e^{-\lambda_{\alpha} |\tau|}\, , $$ from
which we deduce that $\frac{1}{\lambda_{\alpha}}$ is the
correlation time for the $\alpha$-th reservoir.

For different values of the parameter $\lambda_{\alpha}$, this
model interpolates between an ideal Jaynes-Cummings model with
vacuum Rabi frequency $\Omega_{\alpha}$ (which is obtained for
$\lambda_{\alpha} \rightarrow 0$) and a Markovian environment
giving rise to an exponential decay of the atomic excited level
with a rate $\gamma_{\alpha} = 2 \Omega_{\alpha}^2
/\lambda_{\alpha}$ , which is obtained for a very large
$\lambda_{\alpha}$. For intermediate values of the lifetime the
model is known to describe finite memory effects (with a memory
time given by the photon lifetime within the cavity).

Assuming that the two cavities are initially empty and that the
two atoms are prepared in an entangled state, we have \be
\label{eq:stateini} \ket{\psi,0} =  \left (\sqrt{\frac{1-s}{2}}
\ket{g_1g_2} + e^{i \vphi} \sqrt{\frac{1+s}{2}} \ket{e_1e_2}\right
) \bigotimes_{\alpha=1,2} \ket{0_{k_{\alpha}}}\ee where $s \in
[-1,1]$ is a separability parameter, related to the initial tangle
as $\tau_{1,2}(0) =1-s^2$. Notice the the sign of $s$ is
irrelevant for the entanglement, as it only determines the initial
population inversion of the two atoms (that is, $\langle
\sigma_{\alpha}^z(0) \rangle = s$). The time evolution generated
by $H^{(i)}$ gives rise to \be \label{eq:statet} \ket{\psi,t} =
\sqrt{\frac{1-s}{2}} \ket{g_1g_2}\bigotimes_{k_{\alpha}}
\ket{0_{k_{\alpha}}} + e^{i \vphi} \sqrt{\frac{1+s}{2}}
\ket{\phi_1} \otimes \ket{\phi_2},\ee where \be
\label{eq:statephi} \ket{\phi_{\alpha}}= {\cal E}_{\alpha}(t)
\ket{e_{\alpha}} \otimes\ket{0_{k_{\alpha}}} + {\cal
F}_{\alpha}(t) \ket{g_{\alpha}} \otimes
\ket{\mbox{\mbox{photon}}_{\alpha}},\ee with
$\ket{\mbox{\mbox{photon}}_{\alpha}}$ being the (normalized) state
describing the photon emitted in the modes of the $\alpha$-th
reservoir, whose explicit expression is not relevant.

For reservoirs with the spectral density given above, one has:
\begin{eqnarray} && {\cal E}_{\alpha}(t) = e^{-\frac{(\lambda_{\alpha}-i\delta_{\alpha})t}{2}}
\left [\cosh (\frac{w_{\alpha}t}{2} ) + \frac{\lambda_{\alpha}-i
\delta_{\alpha}}{w_{\alpha}} \sinh (\frac{w_{\alpha} t}{2} )
\right ]\, \label{eq:e}, \\ && |{\cal F}_{\alpha}(t)| = \sqrt{1-
|{\cal E}_{\alpha}(t)|^2} \, , \label{eq:f}\end{eqnarray} with
$$w_{\alpha}= \sqrt{\lambda_{\alpha}^2 - 2 i \delta_{\alpha}
\lambda_{\alpha}- 4 R^2_{\alpha}}$$ where
$R_{\alpha}=\sqrt{\Omega_{\alpha}^2 + \delta_{\alpha}^2/4}$.

\subsection{Limiting cases} On resonance, $\delta_{\alpha}=0$, it is
easy to evaluate the limiting expressions for the amplitude ${\cal
E}$. In the Markovian limit, it takes the approximate form ${\cal
E}_{\alpha}(t) = \exp \{ - \gamma_{\alpha}t/2\}$ and the problem
reduces to the one described in \cite{lopez}. On the other hand,
in the Jaynes-Cummings limit one simply has ${\cal E}_{\alpha}(t)
= \cos \Omega_{\alpha} t$.

Some more details about these two limiting cases are reported in
the following subsections.
\subsubsection{Markovian limit}
Taking the limit $\lambda_{\alpha} \rightarrow \infty$, one has
$w_{\alpha} \simeq \lambda_{\alpha} (1-\frac{2
\Omega_{\alpha}^2}{\lambda_{\alpha}^2})$, and furthermore, the
hyperbolic functions can safely be approximated by the positive
exponentials so that
\begin{equation}
{\cal E}_{\alpha} \simeq \exp \{-
\frac{\Omega_{\alpha}^2}{\lambda_{\alpha}} t \} \equiv
e^{-\gamma_{\alpha} t /2}
\end{equation}
This could have been seen directly in the spectral density:
$$\lim_{\lambda_{\alpha} \rightarrow \infty} J_{\alpha}(\omega) =
\lim_{\lambda_{\alpha} \rightarrow \infty} \frac{\gamma_{\alpha}
\lambda_{\alpha}}{2\pi} \, \frac{\lambda_{\alpha}}{(\omega-
\omega_{c_{\alpha}})^2 + \lambda_{\alpha}^2} =
\frac{\gamma_{\alpha}}{2 \pi} \, ,$$ which implies that
$f_{\alpha}(\tau) = \gamma_{\alpha} \, \delta (\tau) \, .$ An
approximately flat spectral density leads to a short memory
(markovian) reservoir. The approximate flatness of
$J_{\alpha}(\omega)$ also leads to the fact that the detuning is
irrelevant as long as it is smaller than the reservoir band-width.
Indeed, by including the leading term in $\delta_{\alpha}$ one has
\begin{equation}
{\cal E}_{\alpha} \simeq \exp \{- \left (
\frac{\delta_{\alpha}^2}{4\lambda_{\alpha}} +
\frac{\gamma_{\alpha}}{2} \right ) t \}
\end{equation}

\subsubsection{Jaynes-Cummings limit}
The ideal cavity limit is obtained for $\lambda_{\alpha}
\rightarrow 0$. By just dropping all of the terms containing
$\lambda_{\alpha}$, one has
\begin{equation} {\cal E}_{\alpha}  =
e^{i \delta_{\alpha} t/2} \Bigl [ \cos R_{\alpha} t - \frac{i
\delta_{\alpha}}{2 R_{\alpha}} \sin R_{\alpha} t \Bigr ]
\end{equation}
Including the leading correction in $\lambda$, the expression for
${\cal E}$ becomes quite cubersome; while it considerably
simplifies in the resonant case ($\delta_{\alpha}=0$) to become
$${\cal E} \simeq e^{-\lambda_{\alpha} t/2} \, \cos \left (
\Omega_{\alpha} - \frac{\lambda_{\alpha}^2}{8 \Omega_{\alpha}}
\right ) t \, .$$

\section{Entanglement dynamics}
\label{sec:3} The general expression for the concurrence as a
function of time is \be C_{1,2}(t) = \mbox{max} \, \left \{ 0 ,
|{\cal E}_1 {\cal E}_2 | \left [ \sqrt{1-s^2} - (1+s) \, \left
|{\cal F}_1 {\cal F}_2 \right | \right ] \right \}\, ,\ee and the
entanglement sudden death is found to occur at a time such that
$$|{\cal F}_1(t) {\cal F}_2(t)| = \sqrt{\frac{1-s}{1+s}} \, .$$
Since $|{\cal F}|\leq 1$, this implies that $s>0$ is required,
which simply indicates that, for ESD to occur, the component of
the initial state taking part to the decay should be larger than
the other, stable one involving the ground states. As noted in
\cite{bellomo}, however, due to the memory of the reservoirs, the
death is (partly) reversible and the entanglement can revive (and
then die again) after some time. In the limit of perfect cavities,
$\lambda \rightarrow 0$, the entanglement dynamics is truly
periodic with a sequence of dark periods followed by perfect
entanglement revivals.

Since the state given above is in the Schmidt form, the reservoir
behave as effective two level systems. This implies that, at any
given time, one can evaluate the entanglement {\it between} the
reservoirs by employing the concurrence again. It is easy to show
that \be C_{r_1,r_2}(t) = \mbox{max} \, \left \{ 0 , |{\cal F}_1
{\cal F}_2 | \left [ \sqrt{1-s^2} - (1+s) \, \left |{\cal E}_1
{\cal E}_2 \right | \right ] \right \}\, ,\ee It is clear from the
form of the initial state that there is no entanglement between
$r_1$ and $r_2$ at the beginning of the time evolution, but such
an entanglement builds up after some time, and a ``sudden birth''
is found to occur.

It also very easy to evaluate the entanglement between qubit
$\alpha$ and its reservoir. Since the reservoirs are effective two
level systems, the concurrence can be used once more, and one
finds: \be C_{\alpha,r_{\alpha}} = (1+s) \left | {\cal E}_{\alpha}
\, {\cal F}_{\alpha} \right | \, .\ee Furthermore, one can use the
tangle to evaluate the entanglement between the two (qubit+cavity)
systems. It is easy to verify that
$$\rho_{\alpha,r_{\alpha}} = \begin{pmatrix}0 & 0 & 0 & 0 \cr 0 &
(1+s) |{\cal E}_{\alpha}|^2/2 & (1+s) {\cal E}_{\alpha} {\cal
F}_{\alpha}^*/2 & 0 \cr 0 & (1+s) {\cal E}_{\alpha}^* {\cal
F}_{\alpha}/2 & (1+s) |{\cal F}_{\alpha}|^2/2 & 0 \cr 0 & 0 & 0 &
(1-s)/2 \end{pmatrix}$$ so that the tangle \be \tau_{(1\otimes
r_1)/(2 \otimes r_2)} \equiv 2 \Bigr ( 1 - \mbox{Tr}
\{\rho_{1,r_1}^2\} \Bigr ) = 1 - s^2 \, ,\ee which is equal to the
square of the initial concurrence between the qubits. Thus, we
have found a simple entanglement invariant: the full Hamiltonian
is bi-local with respect to this bi-partition of the overall
system, and therefore this type of entanglement doesn't change in
time. Notice that this is true irrespective of the form of ${\cal
E}$, so that it holds true for the Jaynes Cummings model
\cite{yudue,mansei} as well as for the markovian decay
\cite{lopez}.

While for entangled states with a single excitation some
conservation law has been found for the squares of the
concurrences, see \cite{yudue,mansei,schan}, for the
two-excitation-case we consider here, only the inequality
$$0 \leq C_{1,2}^2 + C_{r_1,r_2}^2 +C_{1,r_2}^2+C_{2,r_1}^2 \leq \tau_{(1\otimes
r_1)/(2 \otimes r_2)} $$ has been verified, while a more general
relation (not directly involving the concurrences)  has been
reported in \cite{tong}.
\begin{figure}[h]
\includegraphics[scale=.45]{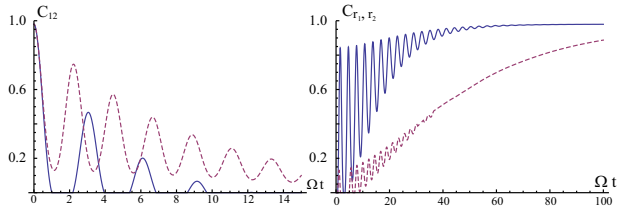}
\caption{Entanglement between the two atoms (left plots) and
between the two reservoirs (right plots) in the good cavity limit
for two equal cavities with $\lambda=0.1 \Omega$ and for two
different detunings: $\delta = \Omega/2$ (solid, blue line) and
$\delta=2 \Omega$ (dashed, purple line). The initial state is such
that $s=0.2$.} \label{figuno}
\end{figure}
As a first example of the time evolution described by the above
expressions for the concurrences, we report in Fig. (\ref{figuno})
the dynamics of atom-atom entanglement, $C_{1,2}(t)$, and
reservoir-reservoir entanglement, $C_{r_1,r_2}(t)$, for two
different detunings and for equal cavities, whose parameters are
very close to the ideal limit (the photon escape rate is chosen to
be ten times smaller the resonant Rabi frequency). Various
features should be noted in these plots; let us concentrate on the
case with smaller detuning first ($\delta = \Omega/2$, solid
lines). Under such conditions, $C_{1,2}$ shows dark periods; that
is, time intervals following an ESD, during which the entanglement
stays equal to zero before abruptly reviving with a periodicity
essentially dictated by the ideal Rabi time, $2 \pi /R$. During
these dark intervals, $C_{r_1,r_2}$ attains its maxima, and the
two concurrences oscillate exactly out of phase. On a time scale
essentially set-up by $\lambda$, both of the bipartite
entanglement measures are seen to become close to their asymptotic
values of $C_{1,2} \rightarrow 0$, and $C_{r_1,r_2} \rightarrow
\sqrt{1-s^2}$ (the saturation value for the inequality reported
above).

One particular aspect that we want to consider here is the effect
of the detuning. It was noticed in Ref. \cite{schan} that, in the
ideal Jaynes-Cummings limit, ESD is quite sensitive to the
detunings. The same statement is clearly seen to hold for the
reservoir entanglement and for the ESB. Indeed, Fig.
(\ref{figuno}) shows that by increasing $\delta$, apart from an
obvious change in the oscillation period, the time evolution is
qualitatively modified as {\it i}) the ESD does not occur anymore
(and, therefore, no dark intervals are found); {\it ii}) it takes
a much longer time before the asymptotic values are approached,
both for the atom-atom and for the reservoir-reservoir
entanglement.

\begin{figure}[h]
\includegraphics[scale=.45]{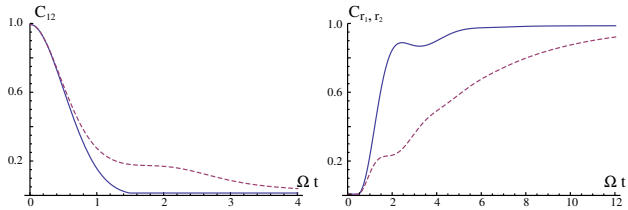}
\caption{Same as Fig. (\ref{figuno}), but with $\lambda = \Omega$.
Here too, the solid, blue plots correspond to $\delta = \Omega/2$;
while the purple, dashed ones correspond to the $\delta=2
\Omega$-case.} \label{gifdue}
\end{figure}

This sensitivity to detunings, however, is seen to be gradually
lost as the cavities become more and more leaky. As an example,
Fig. (\ref{gifdue}) reports the same plot as Fig. (\ref{figuno}),
but for the case $\Omega = \lambda$. It can be seen that, although
the time required to almost reach the stationary values is again
longer for larger detunings, the huge effect of $\delta$ on the
sudden death is not found anymore: the entanglement collapses to
zero in both cases and only the ESD-time is affected by $\delta$.
By further increasing the ratio $\lambda/\Omega$ (which implies
going towards the bad cavity limit), the effect of $\delta$
becomes less and less noticeable.

\begin{figure}[h]
\includegraphics[scale=.45]{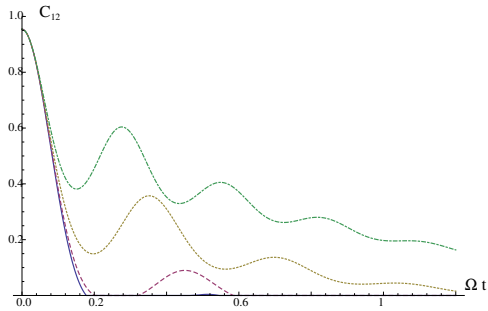}
\caption{Atom-atom entanglement, $C_{1,2}(t)$, evaluated in the
case of equal cavities with $\Omega= 2.5 \lambda$ and with an
atomic initial condition set by $s=0.3$. The four plots (starting
from below) refer to symmetrically detuned atoms with
$\delta_1/\Omega\equiv - \delta_2/\Omega = 0,1,2,3$ (solid,
dashed, dotted and dot-dashed lines, respectively).}
\label{fig:tre}
\end{figure}

A similar behavior is found to occur when the two atoms are
differently detuned from their cavities. The effect is
particularly enhanced when the detunings are opposite in sign
(this is reminiscent of what has been found in Ref. \cite{noi}).
Indeed, by controlling the detuning $\delta=\delta_1=-\delta_2$,
one can qualitatively change the entanglement dynamics and explore
regions in which ESD occurs, others in which dark periods  or
entanglement oscillations are found, eventually reaching parameter
regions in which a monotonic long-time decay takes place. As an
example, in Fig. (\ref{fig:tre}), we report four different
behaviors of the atom-atom entanglement, obtained, for an
intermediate coupling regime with the cavities, by just changing
the detunings.

\section{Entanglement Generation}
\label{sec:4} In this section we describe some schemes for
generating entanglement among one of the atoms and the two
cavities, between the two cavity fields, and between one atom and
a distant cavity. These schemes are based on the assumption that
one can control the couplings between each atom and its
surrounding cavity field as well as the atom-cavity interaction
time. Both these requirements are experimentally feasible, e.g.,
in ion cavity QED systems, since the effective atom-cavity
coupling is achieved via laser driven two-photon processes. The
effective atom-cavity coupling in this case is given by $g_{\rm
eff} = g \Omega /\Delta$, with $g$ the cavity coupling constant,
$\Omega$ the Rabi frequency of the laser driving the Raman
transition, and $\Delta$ the laser and cavity detuning from an
excited third level \cite{harkonen}. In this case, $g_{\rm eff}$
can be controlled by changing the laser intensity and/or detuning
while the interaction time is determined by the length of the
laser pulse.

\subsection{Generation of W-states}
Inserting Eq. (\ref{eq:statephi}) into Eq. (\ref{eq:statet}), we
can explicitly write the total state in terms of its five
components, as follows
\begin{eqnarray}
\ket{\psi,t} &=&
\sqrt{\frac{1-s}{2}} \ket{g_1g_2}\bigotimes_{k_1 k_2}
\ket{0_{k_1} 0_{k_2}} \nonumber \\
&+& e^{i \vphi} \sqrt{\frac{1+s}{2}} \Big[ {\cal E}_1(t) {\cal E}_2(t) \ket{e_1 e_2} \otimes\ket{0_{k_1}0_{k_2}}  \nonumber \\
&+& {\cal E}_1(t) {\cal F}_2(t) \ket{e_1 g_2} \otimes\ket{0_{k_1}, \mbox{\mbox{photon}}_2} \nonumber \\
&+& {\cal F}_1(t) {\cal E}_2(t) \ket{g_1 e_2} \otimes\ket{ \mbox{\mbox{photon}}_1, 0_{k_2}}  \nonumber \\
&+& {\cal F}_1(t) {\cal F}_2(t) \ket{g_1 g_2} \otimes\ket{ \mbox{\mbox{photon}}_1, \mbox{\mbox{photon}}_2} \Big].
\end{eqnarray}
We note that, at a time $\bar{t}$ such that ${\cal F}_1(\bar{t})
\simeq 1$ and ${\cal E}_1(\bar{t}) \simeq 0$, the state of the
system approximates
\begin{eqnarray}
\ket{\psi ' (\bar{t})} &\simeq&\Big\{
\sqrt{\frac{1-s}{2}} \ket{g_2}\bigotimes_{k_1 k_2}
\ket{0_{k_1} 0_{k_2}} \label{eq:W} \\
&+& e^{i \vphi} \sqrt{\frac{1+s}{2}} \Big[   {\cal E}_2(\bar{t}) \ket{e_2} \otimes\ket{ \mbox{\mbox{photon}}_1, 0_{k_2}}  \nonumber \\
&+& (\bar{t}) {\cal F}_2(\bar{t}) \ket{g_2} \otimes\ket{ \mbox{\mbox{photon}}_1, \mbox{\mbox{photon}}_2} \Big]\Big\} \ket{g_1},\nonumber
\end{eqnarray}
From this equation one sees immediately that for $s=1/3$ and
${\cal E}_2(\bar{t}) = \frac{1}{\sqrt{2}}$ one obtains the W-state
\be \ket{\psi_W }= \frac{1}{\sqrt{3}} \left( \ket{010} + \ket{100}
+ \ket{001} \right), \label{eq:Wstate} \ee with
\begin{eqnarray}
\ket{010} &=& \ket{g_2} \bigotimes_{k_1 k_2}
\ket{0_{k_1} 0_{k_2}} , \\
\ket{100} &=& \ket{e_2}  \bigotimes_{ k_2} \ket{ \mbox{\mbox{photon}}_1, 0_{k_2}} , \\
\ket{001} &=& \ket{g_2}  \ket{ \mbox{\mbox{photon}}_1, \mbox{\mbox{photon}}_2}.
\end{eqnarray}
Assuming an intermediate coupling regime for the first cavity,
e.g., $\Omega_1=\lambda_1$, and taking for simplicity
$\delta_1=0$, we see from Eqs. (\ref{eq:e})- (\ref{eq:f})  that
${\cal F}_1(\bar{t}) \approx 0.99 $ at $\lambda_1 \bar{t} = 3$.
The condition ${\cal E}_2(\bar{t}) = \frac{1}{\sqrt{2}}$ is then
obtained for $\Omega_2 \simeq 0.4 \lambda_2$, with
$\lambda_1=\lambda_2$ and $\delta_2=0$.

\subsection{Generation of Bell states}
W-states are characterized by the property that bipartite
entanglement can be obtained after the measurement of any of the
three parties. In our case, measuring the atomic state of the
second atom in $\ket{g_2}$ generates a maximally entangled state
of the cavity fields of the type \be \ket{\psi_{r_1r_2}}=
\frac{1}{\sqrt{2}} \left(   \bigotimes_{k_1 k_2} \ket{0_{k_1}
0_{k_2}} +  \ket{ \mbox{\mbox{photon}}_1, \mbox{\mbox{photon}}_2}
\right). \ee The generation of a W-state followed by a measurement
of the second atom in its ground state is similar to entanglement
swapping, since it amounts at transferring entanglement from the
atoms pairs to two fields inside two distinct non-interacting
cavities.

A Bell state of the atom-cavity system of the second atom can be
generated by measuring the presence of a photon in the other
cavity. In this case the state generated is \be
\ket{\psi_{a_2r_2}}= \frac{1}{\sqrt{2}} \left(   \bigotimes_{k_2}
\ket{e_2 0_{k_2}} + \ket{g_2, \mbox{\mbox{photon}}_2} \right). \ee

Most intriguingly, finally, a measurement of the cavity 2 in the
vacuum creates a maximally entangles state of the atom 2 with the
cavity 1, \be \ket{\psi_{a_2r_1}}= \frac{1}{\sqrt{2}} \left(
\bigotimes_{k_1} \ket{g_2 0_{k_1}} +  \ket{e_2,
\mbox{\mbox{photon}}_1} \right). \ee

Finally we note that a similar W-state can be obtained by
requiring that ${\cal F}_2(t) \simeq 1$ and ${\cal E}_2(t) \simeq
0$. The results are exactly the same as the ones discussed above
provided that one exchanges atom 1 with atom 2.

\section{Summary and conclusions}
\label{sec:5} We have discussed various features related to the
transfer of quantum correlation due  to the interaction of a pair
of atoms (qubit) with two independent bosonic reservoirs, having
non-markovian features. In particular, we have taken a Lorentzian
spectral density for each reservoir, in order to mimic the nearly
resonant coupling of initially entangled atoms to leaky cavities
\cite{manis}. In such a case, the Hamiltonian model is amenable to
an exact solution, so that the time evolution of various bipartite
entanglement measures can be easily evaluated. In particular, we
focussed on  the atom-atom and reservoir-reservoir entanglement to
show that: 1) the initial atomic entanglement is found to be
transferred to the reservoirs; 2) the dynamics of the entanglement
is strongly sensitive to the detuning of the atoms from their
cavities, so that the detuning can be used as a knob to
qualitatively change the character of the time evolution of the
quantum correlations. These conclusions hold true in any
atom-cavity coupling regime, ranging from ideal cavity (strong
coupling regime) with a monochromatic spectral density, to a
markovian environment with an almost structure-less environmental
spectral density (bad cavity or weak coupling limit). Although the
sensitivity to the detuning diminishes by going towards the
markovian limit, we have found that an enhanced sensitivity can be
obtained by independently controlling the two atomic detunings;
and, in particular, that the entanglement time-evolution can be
substantially altered if the two atoms are symmetrically out of
resonance with respect to their cavities.

We have also discussed a protocol to deterministically generate a
W-state with one atom and two cavity fields, which can
subsequently be used as a starting point for probabilistic schemes
aimed at generating maximally entangled states of one atom and
one, distant cavity field, and of the two cavity fields. It is
worth mentioning that these schemes are designed by taking into
account the cavity losses and that they work despite their
presence.

\end{document}